\documentclass[table]{article}
\usepackage[a4paper, margin=1in]{geometry}
\usepackage{authblk}
\usepackage[utf8]{inputenc}
\usepackage[hyphens]{url}
\usepackage{booktabs}
\usepackage{datatool,graphicx,xcolor,cals}
\usepackage{todonotes}
\usepackage{pgfplots}
\usepackage{pgfplotstable}
\usepackage{tikz}
\usepackage{caption}
\usepackage{subcaption}
\usepackage{listings}
\usepackage{balance}
\usepackage{amssymb}

\usepackage{tcolorbox}
\usepackage{listings}
\usepackage{pifont}
\usepackage[inline]{enumitem}
\definecolor{my_orange}{HTML}{F28522}
\definecolor{my_red}{HTML}{FF1F5B}
\definecolor{my_green}{HTML}{00CD6C}

\definecolor{bgcolor}{rgb}{0.95,0.95,0.95}

\usepgfplotslibrary{colorbrewer}
\pgfplotsset{cycle list/Dark2-8}
\pgfplotsset{compat = 1.18} 
\usetikzlibrary{pgfplots.statistics} 

\definecolor{whitesmoke}{rgb}{0.96, 0.96, 0.96}
\definecolor{floralwhite}{rgb}{1.0, 0.98, 0.94}

\providecommand{\keywords}[1]{\textbf{\textit{Keywords: }} #1}

\title{Hello, won't you tell me your name?: Investigating Anonymity Abuse in IPFS}

\author[1]{Christos Karapapas}
\author[1]{Iakovos Pittaras}
\author[1,2]{George C. Polyzos}
\author[3,4]{Constantinos Patsakis}

\affil[1]{Athens University of Economics and Business, Greece}
\affil[2]{School of Data Science, The Chinese University of Hong Kong, Shenzhen, China}
\affil[3]{Athena Research Centre, Greece}
\affil[4]{University of Piraeus}
\affil[ ]{\protect\url{karapapas@aueb.gr}, 
\protect\url{pittaras@aueb.gr},
\protect\url{polyzos@acm.org}, \protect\url{kpatsak@unipi.gr}}
\begin{document}

\date{}
\maketitle
\begin{abstract}
  The InterPlanetary File System~(IPFS) offers a decentralized approach to file storage and sharing, promising resilience and efficiency, while also realizing the Web3 paradigm. Simultaneously, the offered anonymity raises significant questions about potential misuse. In this study, we explore methods that malicious actors can exploit IPFS to upload and disseminate harmful content while remaining anonymous. We evaluate the role of pinning services and public gateways, identifying their capabilities and limitations in maintaining content availability. Using scripts, we systematically test the behavior of these services by uploading malicious files. Our analysis reveals that pinning services and public gateways lack mechanisms to assess or restrict the propagation of malicious content.
  Our findings demonstrate that attackers can exploit the decentralized nature of IPFS and its ecosystem to ensure persistent availability of malicious content while masking their identities. Moreover, we observed instances of this exploitation occurring in practice, further validating the real-world applicability of such attacks.
  
      
  \end{abstract}
\section{Introduction}
Web3, often referred to as the \textit{read-write-own} Web, has recently surged in popularity among users and researchers. Although initially presented as a new phase of the World Wide Web, it primarily represents an ideological shift rather than a technological breakthrough. Its main pillars are decentralization, returning data control to users, and the absence of a central authority, treating all users as peers. To achieve this goal, Web3 engulfs technologies such as blockchains, digital currencies, and decentralized identities, all of which have seen rapid growth.

In terms of security, Web3 seeks to mitigate single points of failure, which, in the recent past, have caused substantial disruptions across various technology sectors, including outages of well-known services, leading to widespread paralysis in different technological domains~\cite{outage}. As with all things in life, Web3 has its dark aspects. The growing interest of users has also drawn the attention of malicious actors toward Web3. Lack of oversight and regulatory authority has led to significant financial losses due to various scams~\cite{bartoletti2021cryptocurrency,patsakis2020unravelling}. On the other hand, as the technologies that comprise Web3 are still in their infancy, they suffer from software vulnerabilities, which are exploited by various actors~\cite{carpentier2023mapping}. Another perspective from which Web3 undeniably faces challenges is that of privacy and anonymity, primarily due to its Peer-to-Peer (P2P) nature. While decentralization is a key advantage of Web3, the inherent transparency and traceability of P2P networks often collide with users' privacy expectations~\cite{sheridan2022web3}.

Web3 consists of multiple stacks, each with various protocols that interoperate to deliver user services. These services range from data storage, domain name resolution, and decentralized identities to applications like social media, gaming, and marketplaces. As these protocols and their interconnections evolve, they present potential vulnerabilities that attackers can exploit or leverage. In data storage, there are various protocols, such as InterPlanetary File System~(IPFS), Filecoin, Storj, SIA, and others.
However, IPFS is widely recognized as one of the most prominent and broadly adopted solutions~\cite{ipfsfriends,trautwein2022design}. Its open-source nature, content-addressable design, and integration with technologies like Filecoin and public gateways have contributed to its popularity across Web3 applications. Developed by Protocol Labs~\cite{protocol} as an open-source project, it has gained considerable attention in recent years. Notably, companies such as Lockheed Martin have shown interest, even launching an IPFS node into orbit~\cite{lockheed}. Furthermore, the growing number of research papers with `IPFS' in their titles highlights its increasing prominence among researchers, with Semantic Scholar returning more than 800 results for such publications over the past two years. Among the tools that enhance the functionality of IPFS are pinning services, which play a crucial role in maintaining file availability across the network. These services allow users to ensure that specific files remain accessible by hosting them on dedicated nodes, even if the original uploader goes offline. Over time, IPFS has drawn the attention of both malicious actors and security researchers. One study~\cite{ipstorm} revealed that a notorious botnet exploited its network, while another highlighted that a significant proportion of its nodes are operated by malicious actors~\cite{ipfsmal}.

Recent works have shown that malware increasingly leverages benign Internet services to distribute payloads and evade detection. This includes both centralized platforms such as GitHub and Dropbox~\cite{yao2023hiding}, and large-scale abuse of cloud services like Discord, Mediafire, and Google Drive~\cite{allegretta2025web}. Our work extends this threat model to decentralized infrastructures like IPFS, where anonymity, content immutability, and the absence of centralized moderation create an even more permissive environment for abuse.
In this paper, we investigate how malicious actors can exploit existing technologies within the IPFS ecosystem to anonymously upload and distribute content. We begin by mapping the current landscape of tools and protocols used to add and access content on IPFS, including pinning services and public gateways. We then design and evaluate practical attacks that leverage these mechanisms to achieve anonymity and persistence within the network. Finally, we explore potential countermeasures to mitigate such exploits. Hence, our main research questions are the following.\begin{enumerate}[label=\bfseries RQ\arabic*:,
    ref=(RQ\arabic*),
    leftmargin=*]
    \item Do pinning services apply the best know your customer (KYC) practices to allow attribution when malicious content is pinned? (Answered in \S\ref{pinattack})\item Do pinning services apply any content scanning mechanism to prevent malicious content sharing? (Answered in \S\ref{pinattack})\item How could an adversary abuse these gaps to share malicious content anonymously? (Answered in \S\ref{pinattack})\item Is there evidence showing this abuse?(Answered in \S\ref{measurements})\item How could an adversary abuse gateways' caching to anonymously share and preserve malicious content online?(Answered in \S\ref{gatewayattack}).
\end{enumerate} 
As a result, our research reveals several ways in which IPFS can be abused without providing the necessary tracking mechanisms for perpetrator attribution. 

\noindent\textbf{Ethical considerations.} While working with live systems, we have taken the necessary measures to ensure that no malware would propagate through the systems and cause any harm. First, any malicious file submitted to IPFS is not executed by any system. However, we acknowledge that once published on IPFS, files may be accessed by third-party systems—including automated scanners or research tools—that might perform dynamic analysis or sandbox execution. To mitigate this risk, all uploaded files either contained benign code flagged as malicious due to simulated behaviors, or were legacy malware samples that no longer pose realistic threats. Secondly, for someone to collect and execute our samples, they must know the CID or monitor all nodes to collect and execute each file. Since the CIDs have not been publicly promoted, the chances of someone collecting all uploaded files and executing them in an unprotected environment are very low. Even in this unlikely scenario, the malicious samples we created just trigger an antivirus without causing actual damage to the system. Finally, the malware we have used from the real world is very well known, and the corresponding URLs have been siphoned, further diminishing the chances of our work impacting any system. We acknowledge that detailing these attack vectors may inadvertently provide insights to malicious actors. However, we believe that openly discussing these issues is necessary to drive improvements in content moderation and security mechanisms within IPFS. Moreover, existing studies have already shown abuse of the IPFS ecosystem. Although no vulnerabilities were directly exploited, we recommend stakeholders in the IPFS ecosystem consider these findings to enhance security measures.

\section{Background}
The \emph{InterPlanetary File System}~(IPFS)~\cite{benet2014ipfs} is a decentralized file-sharing system focusing on distributed data storage and quick file distribution. 
IPFS was created and is
maintained by Protocol Labs as an open-source project
~\cite{protocol}. 
Unlike traditional file systems, IPFS uniquely identifies files based on their content, assigning each file a distinct \emph{Content {ID}entifier}~(CID). A key IPFS component is \emph{libp2p}, an open-source library of network protocols that includes KAD-DHT, a scalable variant of Kademlia \emph{Distributed Hash Table}~(DHT). The KAD-DHT manages three types of mappings, including \emph{Provider Records}, which indicate who hosts specific content; \emph{Peer Records}, which contain information about a specific peer; and \emph{IPNS records}, which link a static address to dynamic data. IPNS names are essentially pointers (IPNS names) to pointers (IPFS CIDs), whereas IPFS CIDs are immutable (because they are derived from the content) pointers to content, Moreover, IPNS names are self-certifying.
\emph{Bitswap}, a key component of IPFS, acts as the data exchange and occasionally as a content discovery protocol, using ``want-have'' and ``have'' messages for efficient data transfer. IPFS employs Merkle DAGs, a combination of Merkle Trees and \emph{Directed Acyclic Graph}~(DAG), to certify the uniqueness of the exchanged data, ensuring that no duplicates are stored. A recent addition to the IPFS ecosystem is the InterPlanetary Network Indexers~(IPNI), a centralized version of the DHT designed to efficiently index provider records. It serves primarily large content providers and complements the existing DHT by focusing solely on provider record management. Additionally, Protocol Labs and other companies offer services that offer public gateways, allowing users to access the content of the IPFS network without maintaining a node.

In IPFS, each peer manages a network of active connections, known as the \emph{swarm}, which typically ranges from $600$ connections~(the \emph{low water mark}) to $900$~(the \emph{high water mark}). When a user requests a file from the IPFS network, the Bitswap protocol is triggered. It sends a message to the user's swarm peers in the format \texttt{want-have <root CID>}~\cite{bitswap}.
Peers in the swarm individually check whether they have the specified CID locally. If a peer possesses the requested content, it responds with a \texttt{have} message. If no response is received within 1~second, the process is handed over to the DHT, which operates in two stages. Initially, the process searches for the Provider Record, which contains the Peer ID, which stores the content for the requested CID. Subsequently, it searches for the Peer Record, which shows how the Peer ID is linked to a network address. Once this process is finalized, Bitswap is reactivated to facilitate data exchange with the peer hosting the content~\cite{trautwein2022design}.

\section{Adding a File to IPFS}\label{adding}
There are several ways to add a file to IPFS. In this section, we explore different methods and their respective modi operandi. 
Additionally, we examine the information about the original uploader that can be retrieved for each method and the duration that the files remain online.

\begin{figure*}[!ht]
    \centering
\includegraphics[width=0.7\textwidth]{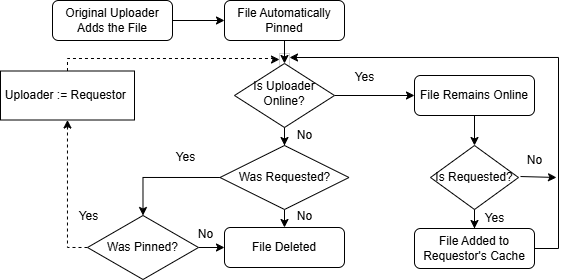}    
    \caption{The File Lifecycle In IPFS.}
    \label{fig:flow}
\end{figure*}

\textbf{IPFS Node}\label{adding_with_node}
For the average user, the primary option for connecting to the IPFS network is the IPFS Desktop application, which supports the most operating systems and includes the functionality of an IPFS node within a user-friendly graphical interface. There is also a command-line version available called Kubo.  When a new node connects to the network, if it has a public IP, it is characterized as a DHT Server. Otherwise, e.g., being behind NAT, it defaults to a DHT Client. This is managed by a mechanism called \emph{Autonat}. This distinction ensures that DHT Servers store and provide data, while DHT Clients only request it, optimizing the network's efficiency~\cite{trautwein2022design}. When a user wishes to publish a file to IPFS, the process involves splitting the original file into smaller chunks, typically $256$~KB. Each chunk is assigned a unique CID and organized into a Merkle DAG added to IPFS. Consequently, two types of records are created in the DHT, with one record stored across a set of 20 specific nodes and the other stored across a different set of 20 nodes. The first type, the Provider Record, indicates who is hosting the file and includes two additional parameters: the republish interval (12 hours by default), which assigns new peers if the original 20 nodes go offline, and the expiration interval (24 hours by default), which verifies that the publisher is still online. The second type is the Peer Record, which maps the peer to its physical address. From the above, it is clear that when a file is added to IPFS, the file itself is not replicated, instead only links pointing to the uploader are created. Upon file addition, the IPFS node automatically pins the file to the original uploader's node, ensuring its availability while the uploader remains online. Replication occurs only if another user requests the file, resulting in it being stored in their cache. Should the original uploader disconnect from the network, the file's availability relies entirely on the cache of interested users. The aforementioned process is depicted in Figure~\ref{fig:flow}.
It is also worth mentioning that the Brave Browser natively supports the use of IPFS in conjunction with a local node~\cite{ipfsbrows}, yet earlier versions provided the ability to add files via Public Gateways.

\textbf{Pinning Services}
IPFS, according to its design principles, does not provide a mechanism to ensure that files added to the network remain online if the original uploader deletes them or disconnects from the network. Files are primarily cached by requesters to ensure their availability to other nodes. The more popular a file is, the higher its chances of staying online for an extended period. Additionally, every IPFS node runs a garbage collector to free up storage space. As a result, cached files are periodically removed, leading some files to disappear from the network over time~\cite{ipfsfriends}. To prevent the garbage collector from removing a file, the user must pin it. Pinning can be categorized into two types: local pinning, where the user configures their node to retain the file, though it will fade once the node disconnects from the network; and remote pinning, where an external provider takes the responsibility to ensure that the file remains pinned~\cite{ipfspin}. 

A plethora of pinning services is available, with Pinata, Filebase, Fleek, and 4EVERLAND being among the most popular. These platforms offer user-friendly graphical interfaces for adding files to the IPFS network, simplifying the process for the average user. Moreover, they provide free storage space for uploading and pinning files, making them accessible to a wide range of users. Once added, the files can be retrieved through public gateways, which act as HTTP access points to the IPFS network.

Although Web3.Storage and NFT.Storage
~\cite{nftstorage}
are not strictly classified as pinning services, their functionality closely resembles traditional pinning solutions, so we include them in this section for completeness. These open-source services, developed by Protocol Labs, are designed to store general and NFT-related data, respectively, in the Web3 era. Both services operate decentralized, leveraging IPFS for content addressing and Filecoin for long-term data preservation rather than offering a pinning service. Web3.Storage is notably free for the community, while NFT.Storage operates under a paid model. NFT.Storage was excluded from further experiments, as it specializes exclusively in NFT metadata storage, which falls outside the scope of our analysis focusing on general-purpose file uploads.
\subsection{Public Gateways}
\label{gateways}
Public gateways act as HTTP entry points to the IPFS network, bridging the Web2 and Web3 ecosystems. They process HTTP requests containing CIDs and relay them to an IPFS node, enabling broader access to the network through conventional Web protocols. Although users cannot directly upload files through a gateway, indirect methods enable this functionality, justifying their classification in this section. Furthermore, the HTTP servers underpinning these gateways leverage caching mechanisms, most commonly the Least Recently Used~(LRU) strategy which optimizes performance and user experience by evicting the least recently accessed content when the cache reaches its capacity~\cite{trautwein2022design}.
Based on the above, it is evident that even if the original uploader disconnects from the IPFS network, the file may remain accessible, cached by gateways, with its persistence primarily influenced by its popularity. During the preparation of this study, we identified $10$ online gateways
~\cite{checker}. 
Using the fingerprinting tool WhatWeb
~\cite{whatweb}, 
we found that nine gateways utilize either Nginx software or Cloudflare proxies, which employ the LRU caching strategy to manage content efficiently.

The fact that public gateways serve as a bridge between the traditional Web and the P2P ecosystem of IPFS makes them very crucial for launching and countering several attacks. For instance, an adversary may host a phishing page on IPFS; however, the content must be rendered from the victim's browser. Thus, the bridge fetches the content from IPFS and brings it to the Web. It must be noted that while there is no official deletion mechanism for IPFS~\cite{politou2020delegated}, some public gateways follow blocking mechanisms to prevent specific content from reaching the Web~\cite{sokoto}. Nevertheless, not all gateways follow the same blocking mechanism and, of course, this does not remove the content from IPFS.

\section{Exploiting IPFS for Anonymity: Attack Scenarios}\label{attack}
The anonymity offered by IPFS can be exploited by malicious actors. In this section, we analyze how attackers leverage methods discussed in Section~\ref{adding} to achieve anonymity, presenting and evaluating two distinct attack scenarios. The code is available at \url{https://github.com/mmlab-aueb/ipfs-anonymity} for reproducing the experiments.

\subsection{The Pinning Service Attack}\label{pinattack}
Pinning services ensure that a file remains online. Therefore, it is logical to consider that an attacker could exploit these services to upload a file and guarantee its availability. However, since our focus is on evaluating the level of anonymity, we first examine the information each pinning service requires from users to allow file uploads, i.e., the Know Your Customer~(KYC) procedure. 
We selected Pinata, Filebase, Fleek, Web3.Storage, and 4EVERLAND based on a systematic Internet search. Specifically, we performed Google queries such as ``top IPFS pinning services'' and ``most popular IPFS pinning services,'' identifying the services most frequently mentioned in developer documentation, technical articles, and community discussions. 
Academic literature specifically evaluating IPFS pinning services remains limited, further justifying the need to consult current developer ecosystems and real-world service availability.
Besides the selected providers, our search also highlighted Infura and Temporal. However, Infura currently restricts access to pre-qualified customers~\cite{metamask}, and Temporal appears to have discontinued operations. Thus, our study focuses exclusively on active and publicly available services, realistically representing the infrastructure accessible to potential anonymous attackers.

The Pinata, Fleek, and Filebase services require an email address for user registration. To achieve higher levels of anonymity, we attempted to use a temporary email service. A temporary email is a disposable email address that allows users to receive emails for a short period, often used to maintain anonymity or avoid spam during registration processes. During December 2024 and January 2025, we tested the registration process on Pinata, Fleek, and Filebase using email addresses generated by the service TempMail (\url{https://temp-mail.org}). Both Pinata and Fleek accepted the first temporary email we generated, allowing us to create accounts successfully. After four attempts with different temporary email addresses, Filebase accepted the registration, suggesting that its filtering against disposable emails may be incomplete. In all three cases, the platforms required us to verify the email address using a one-time password (OTP).
4EVERLAND, on the other hand, does not use email-based registration but instead requires a cryptocurrency wallet. Using Metamask, we successfully created an account on the platform, noting that even for creating the Metamask wallet, no email was needed. Finally, while Web3.storage accepted the temporary registration email, uploading files required linking a payment account, even though the platform also offers a free plan. This suggests that, although temporary emails are allowed, the payment account requirement serves as an additional verification step for users, limiting its suitability for fully anonymous abuse scenarios. Table~\ref{tab:kyc} presents a summary of these findings.

\begin{table*}[t]\label{table:kyc}
\centering
\scriptsize
\caption{Registration Requirements \& Free Storage for Pinning Services.}
\label{tab:kyc}
\begin{tabular}{|l|l|c|c|c|c|c|}
\hline
\textbf{Pinning} & \textbf{URL} & \textbf{KYC} & \textbf{Temp Mail} & \textbf{Free} & \textbf{Registered}& \textbf{DMCA}\\ 
\textbf{Service} &  & & \textbf{Accepted} & \textbf{Storage} & \textbf{Country} & \textbf{Compliant}\\ 
\hline
Pinata     &\url{https://pinata.cloud}     & E-mail    &         \checkmark           &        1 GB & USA   & \checkmark   \\ \hline
Filebase        & \url{https://filebase.com} & E-mail  &         \checkmark           &   5 GB & USA   & \checkmark     \\ \hline
Fleek  & \url{https://fleek.co}         & E-mail     &   \checkmark                 &  5 GB & USA & \checkmark         \\ \hline
Web3.Storage    & \url{https://web3.storage} & Credit Card  &     \checkmark               &      5 GB & USA   & -     \\ \hline
4EVERLAND     & \url{https://4everland.org} &  Crypto Wallet  &         N/A           &          5 GB & AUS  & \checkmark  \\ \hline
\end{tabular}
\end{table*}

To simulate malicious behavior, we developed a Python script packaged into a Windows executable using PyInstaller~\cite{pyinstaller}. It mimicked keylogging, dummy process injection, basic file manipulation, and failed network connections. The file was safe by design, yet flagged by multiple antivirus engines on VirusTotal~\cite{virustotal} due to behavioral heuristics. No harmful payload or external communication was included. To ensure unique Content Identifiers~(CIDs), we created a distinct version of each script for each pinning service under evaluation. One of the key questions explored in this section is how pinning services handle files clearly marked as malicious, aiming to better replicate the perspective and actions of a potential attacker.
In addition to the simulated malware, we also tested uploading known deprecated malware, specifically the WannaCry ransomware, to the pinning services. The result was identical: the file was successfully uploaded, and its CID was generated. Furthermore, we confirmed its accessibility through the public gateways.
Notably, all files, including WannaCry, were immediately accessible, highlighting the absence of mechanisms in public gateways to evaluate the maliciousness of uploaded content. This raises significant concerns about the potential misuse of the IPFS network.

As previously discussed, in IPFS, the physical address of the node hosting a file can be identified. However, when files are hosted by pinning services, attackers are not concerned about their own address being exposed. The only potential exposure point is during the interaction with the pinning service's website for registration and file upload. To mitigate this risk, an attacker could use a public network or leverage the Tor~\cite{dingledine2004tor} network to enhance their anonymity prior to registering and uploading files to the pinning services. Since many services implement protections that restrict access via Tor, we conducted a series of tests to verify the feasibility of using Tor to access these services. Our tests confirmed that files could be successfully uploaded, and the recorded IP address differed from our actual address, ensuring the attacker's anonymity.

It is important to note that visitors to these files, once uploaded by the attacker, may include either unsuspecting users who were targeted by phishing~\cite{sokoto} or malware campaigns, or, in CyberCrime-as-a-Service scenarios~\cite{raas}, collaborators of the attacker, such as affiliates. Even in the latter case, leveraging the Tor network can effectively mitigate the risk of exposing their identities or the nature of their activities.

\begin{figure*}[!ht]
    \centering
    \includegraphics[width=0.7\textwidth]{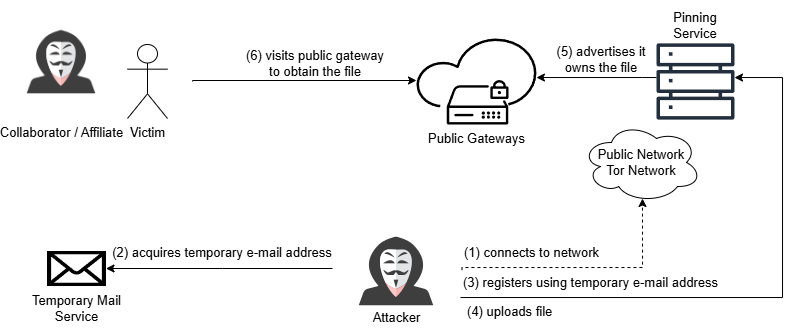}
    \caption {Design of the ``Pinning Service Attack''.}
    \label{fig:attack1}
\end{figure*}

Figure~\ref{fig:attack1} presents the steps that a malicious actor must follow to execute the ``Pinning Service Attack''. It allows the attacker to leverage the Tor network for anonymity and anonymously upload files to IPFS. By utilizing pinning services, the attacker ensures that uploaded files remain persistently online.

\subsection{The Public Gateway Attack}\label{gatewayattack}

As mentioned, Public Gateways of IPFS do not provide a direct method for uploading a file to the network. However, their caching might indirectly serve as a pinning service, providing file availability. In this section, we initially examine whether and for how long a file remains cached.

To better understand this phenomenon, we conducted a systematic experiment focusing on caching behavior across multiple gateways. The methodology we adopted is as follows. From the 10 gateways identified in Section~\ref{gateways}, we selected five based on their strong association with well-known Web companies (e.g., Pinata, Infura), official status within the IPFS ecosystem, reputation, and service quality. Specifically, we chose \begin{enumerate*}[label=(\alph*)]\item \url{ipfs.io} (the official gateway maintained by Protocol Labs), \item \url{gateway.pinata.cloud}, \item \url{infura-ipfs.io}, \item \url{flk-ipfs.xyz} and \item\url{4everland.io}\end{enumerate*}. For each selected gateway, we created four different files, resulting in 20 different files. First, we wanted each gateway to have different files to avoid cross-caching scenarios. Second, for each of these, we created four different files corresponding to the four time scenarios we are studying: 1 hour, 6 hours, 12 hours, and 24 hours. We use these intervals to request the respective files from the gateways to understand how popular a file needs to be to remain cached. Subsequently, we used an IPFS node to add the files, ensuring our node ran as a DHT server. Then, to confirm that all the gateways cached all files, we sent up to four requests per file to verify their caching status. The four requests were performed in a negligible amount of time, less than five minutes, and the files became available. After successfully ensuring that all files were cached across the gateways, we disconnected the node from the network, leaving the gateways as the sole source of file hosting. The latter allows us to isolate the role of gateway caching in maintaining file availability independent of the original node. By doing so, we could analyze how the caching mechanisms of public gateways sustain file accessibility over time.

\begin{figure*}[!ht]
    \centering
    \includegraphics[width=.9\textwidth]{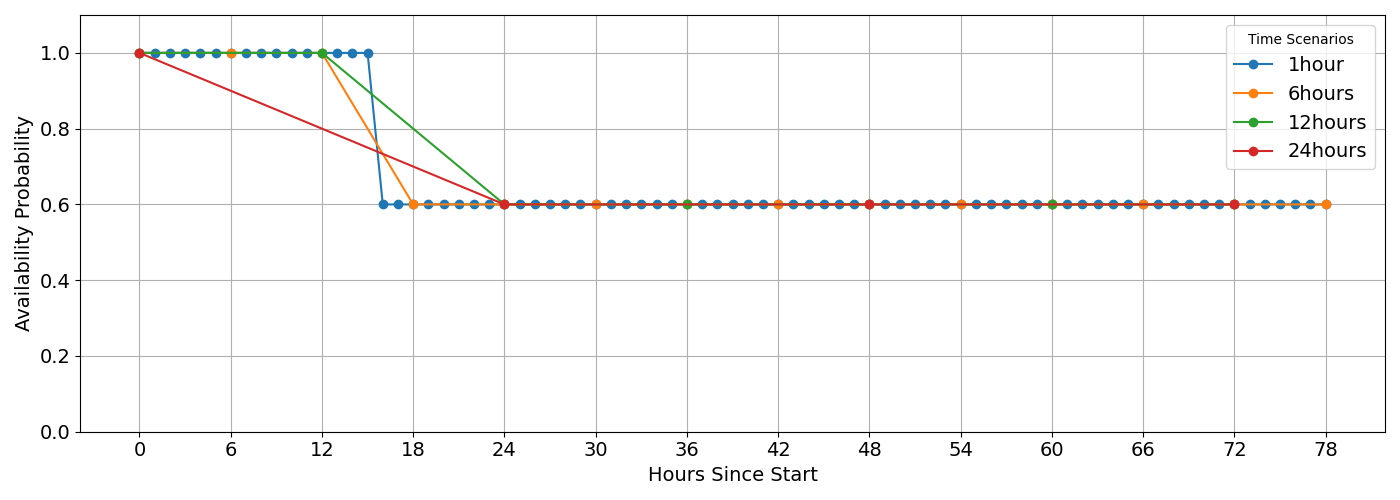}
    \caption {Time-Dependent File Availability Analysis.}
    \label{fig:availability}
\end{figure*}

We automated the process of sending requests to the gateways based on the aforementioned periods and recorded the responses for more than three days. The results indicate that caching duration varies significantly between gateways, with some maintaining availability longer than others, which could be attributed to differences in caching strategies or the relative popularity of each gateway. Figure \ref{fig:availability} illustrates the ratio $\frac{\nu}{5}$ per hour, where $\nu$ represents the number of gateways caching our files at a given time across the different time scenarios. As depicted, two out of the five gateways removed our files from their cache shortly after 16 hours, while the remaining three continued to retain them online. For ethical reasons, we refrain from disclosing which ones retained or removed the files.

In conclusion, we have demonstrated that a malicious actor could potentially exploit Public Gateways to maintain files on the IPFS network anonymously. The process involves first uploading the files to the IPFS network and generating artificial traffic by repeatedly requesting these files. This ensures that the Public Gateways cache the files. Once the files are cached, the actor can sustain their availability by periodically sending requests for the files, preventing them from being removed from the cache due to inactivity. This approach allows the actor to leverage the distributed infrastructure of Public Gateways to maintain file availability while preserving anonymity, eliminating the need for a dedicated pinning service.
\begin{figure*}[!ht]
    \centering
    \includegraphics[width=0.7\textwidth]{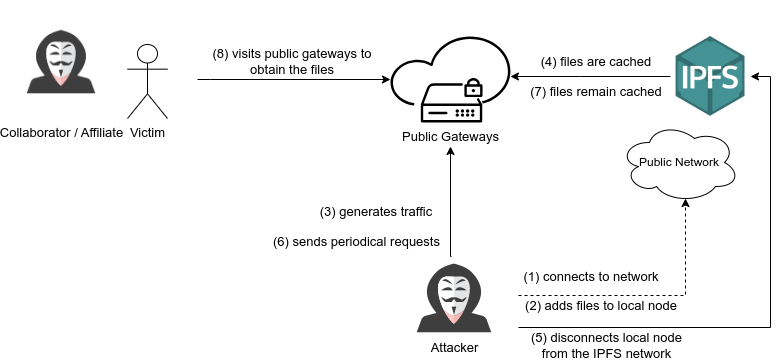}
    \caption {Design of the ``Public Gateway Attack''.}
    \label{fig:attack2}
\end{figure*}
At this point, it should be noted that during the attack, the attacker only risks revealing their physical address while uploading the files via the local node. As previously mentioned, this process requires minimal time, significantly reducing the exposure window for the attacker. Additionally, the attacker could perform this step through a public network to further obscure their physical location. The subsequent periodic requests to the public gateways can also be accomplished through a public network or Tor. Additionally, the attacker could utilize a botnet under their control to generate artificial traffic towards the files without revealing their identity. By distributing requests across multiple geographically dispersed nodes, the botnet obscures the origin of the traffic, making it significantly harder to trace back to the attacker. Note that in the past, the IPFS network has been a victim of such botnet activity~\cite{ipstorm}. A step-by-step implementation of the attack is illustrated in Figure~\ref{fig:attack2}. 

\subsection{Double Extortion Attack}
Typically, ransomware attacks encrypt the victim's files and demand a ransom to be paid to hand over the decryption key. Nevertheless, modern organizations have invested in backup systems that limit the damages of a potential ransomware attack, significantly decreasing the amount of ransom they would be willing to pay. As a countermeasure, ransomware gangs siphon sensitive data to their premises, threatening their victims by leaking the data and creating what is often called a ``double extortion''.

\begin{figure*}[h]
    \centering
    \includegraphics[width=0.7\textwidth]{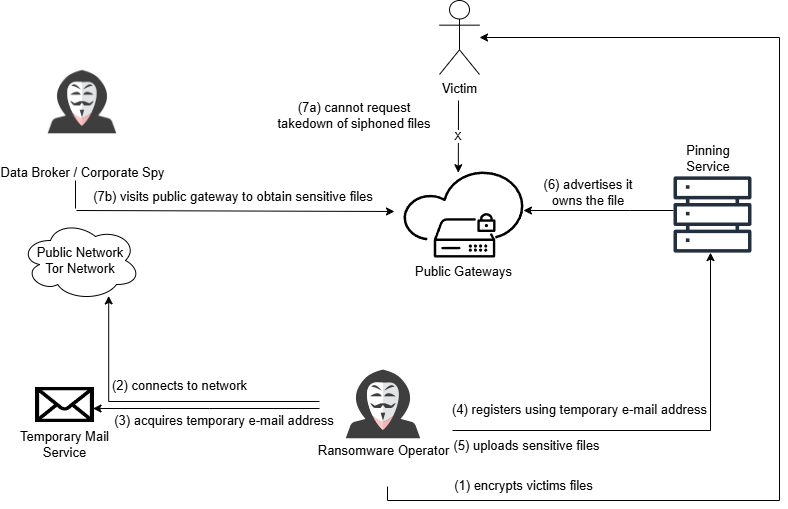}
    \caption {Design of the ``Double Extortion Attack''.}
    \label{fig:attack3}
\end{figure*}

The siphoning of the data can be performed in multiple ways, however, methods like DNS tunneling, while effective, can be very slow. Therefore, ransomware gangs tend to abuse cloud service providers to upload their ``loot''. For example, the notorious Conti group used RClone to upload data to multiple cloud storage providers \cite{sophos}. With IPFS and the poor KYC practices of pinning services, ransomware gangs can have another more robust option. They may harvest sensitive information from the infected hosts and upload them to IPFS through pinning services. Beyond exploiting KYC to gain the necessary storage, ransomware gangs may also exploit whitelisted domains and the lack of content takedown mechanisms. Note that cloud service providers respond to takedown notices, e.g., the victim notifies the cloud service provider that leaked sensitive data are hosted and must be taken down. However, pinning services cannot remove content from the IPFS once it has been uploaded. Although pinning services comply with DMCA policies and can remove a pinned file from their hosted storage, this does not translate into the deletion of the file from the IPFS network. The decentralized nature of IPFS makes this nearly impossible, while the existence of public gateways, many of which do not adhere to the badbits list (as mentioned in \ref{measurements}), further complicates takedown efforts.
Figure \ref{fig:attack3} illustrates this abuse scenario.

\subsection{Real-Life Evidence of Malicious Exploitation}\label{measurements}
While previous work such as~\cite{sokoto} investigated the presence of malicious or illegal content across the IPFS network, our approach specifically targets pinning services, i.e., entities that intentionally maintain long-term availability of hosted content. By focusing on CIDs advertised by major pinning providers, our analysis offers a more precise view into deliberate, persistent misuse of the IPFS ecosystem and links it directly to infrastructures that facilitate anonymity and permanence.

We utilized \texttt{ipni-cli}
~\cite{ipni} 
to monitor CIDs advertised by Pinata, Filebase, and Fleek pinning services on the \texttt{\url{cid.contact}} indexer for 24 hours.
For all providers, we repeatedly executed the following command:
\begin{verbatim}
ipni ads get --ai=<provider addr> --head
\end{verbatim}
This command retrieves information about the latest advertisement from the specified provider, including the number of CIDs it contains.
Once we obtained this information, we proceeded to extract the actual CIDs using: 
\begin{verbatim}
ipni random <provider addr>
\end{verbatim}
With the parameter \texttt{n}, this command returns \texttt{m} CIDs from a random selection of the most recent \texttt{n} advertisements. By setting \texttt{n=1}, we ensured that the selection always targeted the most recent advertisement. Since the previous command had already provided us with the exact number of CIDs, we could request all of them at once. This approach enabled us to systematically retrieve all hashes from every advertisement recorded since the beginning of the experiment. By continuously executing these queries and storing the results, we effectively built a historical record of all advertisements and their associated CIDs from each provider.
During the 24-hour interval, we collected 
\begin{enumerate*}[label=\bfseries(\roman*),
    ref=(\roman*),
    leftmargin=*,
     labelsep=1em  ]
    \item $1,124,780$ CIDs from Pinata,
    \item $718,578$ from Filebase, and
    \item $339,684$ from Fleek
\end{enumerate*}. For each of these, we standardized the format of the CIDs to match the entries in the \texttt{Bad Bits Denylist}
~\cite{badbits}, 
ensuring compatibility for an accurate comparison. The Bad Bits Denylist is a list maintained by Protocol Labs, updated upon email recommendations to filter undesirable files, such as malware, phishing content, or copyright-infringing materials. Note that the list is enforced on the public gateways operated by Protocol Labs but is advisory for all other nodes within the IPFS network. By matching the monitored CIDs against the entries in the denylist, we discovered that within 24 hours, the pinning services advertised five CIDs included in the Bad Bits Denylist. 
It is worth mentioning that one of these CIDs was advertised by all three services, while two were common to two services. We consider the presence of these blocked CIDs --and even more so their simultaneous advertisement on the same day by multiple pinning services-- a strong indication of malicious actors' organized exploitation of the anonymity provided by pinning services. Finally, we managed to retrieve three of them, discovering that one was a JavaScript file involved in a Bank of America phishing scam, the second was a login phishing webpage targeting a Korean webmail service, and the third was an image, likely used for malicious purposes.

\section{Related Work}
Research has shown that malware increasingly abuses centralized Web and cloud platforms for infrastructure, persistence, and evasion. Yao et al.~\cite{yao2023hiding} propose Marsea, a concolic execution engine that detects malware interaction with benign Web applications such as GitHub and Dropbox, revealing how these services are repurposed for malicious use. At a broader scale, Allegretta et al.~\cite{allegretta2025web} analyze threat intelligence from 36 vendors and identify over 22,000 abused benign domains, including services like Discord and Google Drive, used to distribute malware. These works demonstrate that even trusted, centrally managed services are vulnerable to abuse. In this work, we show that decentralized infrastructures like IPFS introduce new and arguably more permissive abuse surfaces, due to their inherent anonymity, lack of content moderation, and resistance to takedown.

In recent years, Web3 has emerged as a new paradigm for the Internet, prioritizing user anonymity and privacy. These features are especially significant as concerns about user privacy and tracking escalate. However, numerous studies indicate that these features are often compromised.  
Kshetri~\cite{kshetri2023} highlights several vulnerabilities within Web3 and the metaverse, particularly the extensive data collection and exposure of personal and sensitive data due to numerous security breaches on Web3. Furthermore, the author points out that anonymity can be compromised via the traceability of blockchain transactions on Web3 platforms, potentially linking personal identities and actions to public transaction records. 
 
On the other hand, other studies focus on how anonymity and privacy are compromised on Web3. 
Wang et al.~\cite{wang2024} explore how Web3 social platforms, such as friend.tech
~\cite{friend}, 
impact user privacy and anonymity. In particular, they identified that the integration between Web3 and legacy Web2 platforms could significantly undermine Web3 anonymity and lead to privacy leakage. This occurs because user actions on Web2 platforms can be associated with accounts on Web3 platforms since these actions are immutably written on blockchains. Then, the recorded actions can be linked and traced back to the users. To address these problems, the authors argue that a balanced approach between transparency and privacy in Web3 is needed. 
Additionally, Torres et al.~\cite{torres2023} focus on how wallets and Decentralized Applications~(DApps) manage user data. The authors conclude that current privacy measures are insufficient, highlighting that Web3 applications, particularly wallets, often expose sensitive user data, such as wallet addresses. This exposure directly contradicts the foundational privacy promises of Web3 by compromising user anonymity and privacy. 

A central element of Web3 and a core focus of our study are distributed file systems, with IPFS being the most prominent. Previous research has demonstrated that IPFS can be exploited by malicious actors in various domains. For example, studies have shown its use in Malware as a Service systems~\cite{raas}, while others have reported the presence of phishing files or copyright violations within the IPFS network~\cite{sokoto}. Moreover, IPFS also has some privacy violations. 
In particular, Balduf et al.~\cite{balduf2022monitoring} showcase a privacy attack on the IPFS network by leveraging the Bitswap protocol and introducing a set of attack vectors. The authors state that every IPFS node is susceptible to each of the introduced attacks, and moreover, they succeed in exploiting it by deploying a number of nodes with extended connectivity to passively monitor the Bitswap channel and demonstrate their attack methodology by discovering the PeerId of the public IPFS HTTP gateways.  

In addition to attackers, security analysts can leverage Bitswap's privacy shortcomings. Son et al.~\cite{forensics} propose \emph{IF-DSS}, a digital forensics investigation framework for Decentralized Storage Services~(DSSs). They analyze the most critical DSSs from the point of view of digital forensics and apply the proposed framework to IPFS. To collect appropriate and sufficient data, they separate them into those that exist on the local side as well as remotely. Finally, they suggest tackling the dissemination of illegal material in three steps: \begin{enumerate*}[label=(\roman*)] \item Content filtering, i.e., blacklisting of the inappropriate content, \item stop content sharing, i.e., turn the node from server to client, and finally, \item shutting down the node.
\end{enumerate*}

On the other hand, some works try to enhance IPFS privacy. 
Katsantas et al.~\cite{katsantas2024} focus on hiding the identity of content on IPFS by using only hash functions. The authors aim to prevent intermediaries from detecting the retrieved contents without relying on trusted third parties. 
Furthermore, Daniel et al.~\cite{bitswap_priv} point out that as IPFS follows the ICN paradigm, a client requests content directly rather than visiting an address. Thus, Bitswap queries all the client's neighbors for content, resulting in the client's interest leaking. Aiming to reduce interest leakage, the authors propose three privacy-enhanced standards for content discovery. By using these protocols, on the one hand, the level of privacy of the client is improved, but that of the provider is reduced. More specifically, they propose a solution using bloom filters and \texttt{Bloom-Swap}, a solution using bloom filters in which the provider sends its inventory to the client, and he, in turn, checks locally whether the requested content is a Bloom Filter member to ask the block directly. \texttt{PSI-Swap}, which uses Private Set Intersection (PSI), reduces and improves privacy levels on the provider's side as well. 
Finally, the \texttt{BEPSI-Swap}, which combines the two previous ones, improves the efficiency of PSI-Swap, at the cost of making PSI probabilistic. 
The authors then implement a proof of concept of the proposed protocols and study them from the security and efficiency perspectives.

\section{Countermeasures \& Conclusions}
The decentralized nature of the technologies we study, combined with the fact that the majority of the software is open-source, makes enforcing rules for implementing countermeasures challenging.
From the perspective of pinning services, KYC practices must become stricter. Measures such as filtering temporary emails, implementing blockchain-based identity systems, e.g., cryptocurrency wallets with benign transaction history, applying stricter criteria for users operating through Tor networks, enabling content scanning mechanisms, and adhering to a centralized deny list like Bad Bits should be enforced. Public gateways act as bridges for Web2 users to access the Web3 ecosystem. For the average user, requiring a blockchain-based identity would deter them from utilizing these gateways. However, all gateways could be required to comply with the Bad Bits, a policy currently enforced only on gateways managed by Protocol Labs. Moreover, even if a CID is listed on the Bad Bits Denylist, a malicious actor can circumvent it by simply choosing an alternative chunking size when adding the file to IPFS~(\textbf{RQ5}). This approach generates a different CID that is not associated with the blacklisted one~\cite{sokoto}, making content filtering on gateways significantly more challenging. In this study, we examined the vulnerabilities of IPFS pinning services and public gateways, highlighting how malicious actors can exploit their anonymity features or lack of proper KYC policies to share undesirable content. By implementing and testing two distinct attack methodologies, we demonstrated not only their feasibility~(\textbf{RQ3}) but also observed instances of malicious activity occurring within the IPFS ecosystem~(\textbf{RQ4}). Our findings reveal critical issues, including the lack of robust KYC practices in pinning services~(\textbf{RQ1}), insufficient content filtering mechanisms~(\textbf{RQ2}), and the challenges posed by the decentralized and open-source nature of the IPFS ecosystem. These gaps enable attackers to take advantage of the anonymity features of the system while avoiding accountability. Since current KYC practices in pinning services can be easily bypassed, the use of stricter measures, of even the consideration of blockchain-based identity verification methods, such as zero-knowledge proofs (ZKPs), e.g., zkLogin \cite{3658644.3690356}, would allow users to verify their legitimacy without exposing their full identity.

It should be stressed that the decentralized nature of IPFS raises significant legal and regulatory challenges, particularly in the enforcement of content moderation and compliance with existing digital laws. While platforms operating in centralized environments are bound by regulations such as the Digital Services Act (DSA)~
\cite{dsa}, 
decentralized systems like IPFS lack clear accountability structures. This creates a regulatory gap that malicious actors can exploit to distribute illicit content while avoiding legal repercussions.
One of the main concerns is jurisdictional ambiguity. Since IPFS content is hosted on a distributed network of peers, it is often unclear which jurisdiction has the authority to enforce takedown requests or prosecute offenders. This is especially true on platforms like IPFS, where there is no deletion mechanism and data ownership is not always known. Pinning services, many of which operate in different countries with varying legal requirements, further complicate the enforcement process.

However, this sparks the debate surrounding IPFS security and other such platforms regarding the trade-off between privacy and censorship resistance. While decentralization offers increased resilience against state-sponsored censorship, it also enables unmoderated content proliferation, including, but not limited to, extremist propaganda, child sexual abuse material, and malware distribution. The ability of malicious actors to exploit anonymity for illegal activities creates a dilemma where content moderation mechanisms must be introduced without undermining the fundamental principles of decentralized storage. Strengthening the security of IPFS and the surrounding ecosystem is essential not only to prevent its misuse but also to promote its adoption as a reliable and privacy-preserving tool for decentralized file sharing, which is fundamental to the Web3 paradigm. Thus, future research could focus on the development of automated tools to detect malicious CIDs in a decentralized and scalable way. Another approach would be decentralized content moderation, where community-driven flagging mechanisms allow for voluntary filtering rather than direct deletion. Likewise, user-driven reputation systems for pinning services and nodes could help differentiate legitimate operators from malicious ones. By assigning trust scores to nodes based on their activity and compliance with community standards, users could make informed choices about which nodes to trust for content retrieval and caching.

\section*{Acknowledgements}
This work was supported in part 
by the European Commission under the Horizon Europe Programme, as part of the project SafeHorizon (Grant Agreement no. 101168562). 
The content of this article does not reflect the official opinion of the European Union. Responsibility for the information and views expressed therein lies entirely with the authors.

    \end{document}